\documentstyle[12pt,epsfig]{article}

\def\be{\begin{eqnarray}}
\def\ee{\end{eqnarray}}
\def\nn{\nonumber}
\def\ds{\displaystyle}
\def\de{\partial}

\def\ob{{\bar\omega}}
\def\L{\mbox{L}}
\def\H{\mbox{H}}
\def\t{\mbox{t}}
\def\lLb{[\![}
\def\rLb{]\!]}
\def\lPb{ \{\!\!\!\!\{ }
\def\rPb{ \}\!\!\!\!\} }
\def\BlPb{ \Big\{\!\!\!\!\Big\{ }
\def\BrPb{ \Big\}\!\!\!\!\Big\} }
\def\p{\mbox{\tt p}}
\def\pp{\mbox{{\scriptsize\tt p}}}

\begin{document}
\begin{flushright}
{\large{\sl UPRF-96-462}}
\end{flushright}
\vskip1.5cm
\begin{center}
{\LARGE  Dynamics as Shadow of Phase Space Geometry}\\
\vskip1.5cm
{\Large J.\ R.\ Klauder$^\dagger$ and P.\ Maraner$^\ddagger$}\\
\medskip
{\large\sl $\dagger$ Departments of Physics and Mathematics,}\\
\smallskip
{\large\sl University of Florida, Gainesville, FL 32611, USA}\\
\medskip
{\large\sl $\ddagger$ Dipartimento di Fisica, Universit\`a di Parma,}\\
\smallskip
{\large\sl and INFN, Gruppo collegato di Parma,}\\
\smallskip
{\large\sl  Viale delle Scienze, 43100 Parma, Italy}
\end{center}
\bigskip 
\begin{abstract}
  Starting with the generally well accepted opinion that quantizing
an arbitrary Hamiltonian system involves picking out some additional
structure on the classical phase space (the {\sl shadow} of quantum
mechanics in the classical theory), we describe classical as well
as quantum dynamics as a purely geometrical effect by introducing a
{\sl phase space metric structure}.
This produces an ${\cal O}(\hbar)$ modification of the classical 
equations of motion reducing at the same time the quantization of an 
arbitrary Hamiltonian system to standard procedures. Our analysis is 
carried out in analogy with the adiabatic motion of a charged 
particle in a curved background (the additional metric structure) under 
the influence of a universal magnetic field (the classical symplectic
structure).
This allows one to picture dynamics in an unusual way, and
reveals a dynamical mechanism that produces the selection of the
right set of physical quantum states.
\end{abstract}

\newpage
\
\vskip1.0cm
\noindent
{\large\sl Dynamics as Geometry }
 
\vskip1.0cm
\noindent
{\large\tt Paolo Maraner}\\
{\large\tt Universit\`a di Parma}\\
{\large\tt Dipartimento Fisico}\\
{\large\tt Viale delle Scienze}\\
{\large\tt 43100 Parma}\\
{\large\tt Italy}

\vskip1.0cm
\noindent
{\large\tt Telephone: +39-521-905281}\\
{\large\tt Fax\ \ \ \ \ \ : +39-521-905223}\\
{\large{\tt E-Mail :}{\sl maraner@parma.infn.it}}
\newpage

\section{Introduction}

 The search for a {\sl geometrical} description of the laws of nature
is part of an important tradition in modern physics, and such techniques
allow one to gain a coordinate-free viewpoint 
of the formal structures as well as the global features of a theory.
 Guided by this spirit, we would like to show how classical (Hamiltonian)
as well as quantum dynamics may be formulated as the {\sl adiabatic limit}
of a {\sl fully-geometrical phase-space-theory} constructed with the help
of a metric and the standard symplectic \cite{A&M,Arn}
phase space structures (see Eq.\ref{mHvp} and Eqs.\ref{ourP},\ref{ourH} 
below). 
 Our work is mainly motivated by the attempt to overcome
various difficulties concerning the construction of a coordinate-free 
quantization procedure, a context in which the introduction of 
subsidiary geometrical phase space structures seems to be unavoidable.

 As is well known, the standard way to look at quantization proceeds from  
Dirac's observation \cite{Dir} that the necessity of interpreting every 
quantum  phenomenon with classical expressions forces the formal structure of
a quantum theory to be isomorphic to the one of the corresponding classical
theory (correspondence principle).
 Therefore, quantization may be regarded as the 
attempt of building a bridge between the formal structures of classical 
and quantum mechanics, that is to say, to find a 
correspondence between classical and quantum {\sl states}, 
{\sl observables  (kinematics)} and {\sl evolution equations (dynamics)}.
 The conclusion emerging from many attempts at building 
 a geometrical quantization procedure
is that it is impossible 
to have a one-to-one correspondence between the algebra of classical and 
quantum observables without making the Hilbert space of the corresponding 
quantum system too large. Moreover the necessary selection of a subalgebra
for which the correspondence holds may be
regarded as picking out some {\em additional
structure} on the classical phase space $M$. {\em ``This {\rm [additional structure]} can be thought
as the {\bf shadow} of quantum mechanics in the classical system and
the element of choice in this selection is the (...) point at which we 
come across an ambiguity in passing from the classical to the quantum
domain''} (N.\ Woodhouse \cite{Woo}, emphasis added).

 The nature of the additional structure is today a matter of discussion. 
The first concerted effort to overcome the difficulty goes back to the first 
Geometric Quantization papers \cite{Kos,Kir,Sou}. It consists in picking out a 
{\sl real} or a {\sl complex polarization} on the phase space $M$---when 
there is one---and asserting that the physical states of the quantum system 
should in some sense preserve the polarization (see e.g.\ \cite{Woo}).
 This prescription emerges from the analysis of a 
wide class of examples (with a high degree of symmetry).
It gives correct physical answers for highly symmetrical 
systems \cite{Ono} but appears problematic as soon as the 
dynamics of systems with less symmetry or no symmetry at all is considered.
There is no longer any guarantee that the evolution of the system respects the 
polarization, and physical states may evolve into non-physical ones.

 Additionally, physical insight into the problem may be gained by looking at 
the phase space path integral expression of the propagator
\be
{ K}(q'',t'';q',t')\stackrel{\mbox{\scriptsize ?}}{=}
       \int
       e^{{i\over\hbar}\int [p_\mu{\dot q}^{\mu}-h(q,p)] dt}\,
       {\cal D}q {\cal D}p.
\label{FP}
\ee
This formal integral involves only the classical symplectic structure and
superficially appears covariant under canonical transformations. It is on 
the  other hand  immediate that this canonical invariance
must be broken. Otherwise the introduction of a suitable set of canonical
variables would make the formal path integral expressions cooincide
and hence make the spectra of distinct physical systems equal.
This undesirable consequence is avoided when it is recognized that to be 
defined the formal integral needs {\em regularization} and that 
regularization---e.g.\ the commonly used lattice regularization---breaks 
canonical invariance. {\em It is the phase space structure producing 
the breakdown of canonical invariance in the conventioanl
phase space path integral 
that can be identified with the {\bf shadow} of quantum mechanics 
in the classical theory.}
 Being restricted to flat phase spaces---moreover to Cartesian
coordinate frames---we cannot hope to gain real insight into the nature of 
the ``additional structure'' by considering  lattice regularizations. In so
doing one must confront the meaning of the formal expression ${\cal D}q{\cal
D}p$, which on the surface appears to be a construct solely of the symplectic
 structure. Nevertheless, for a $q$ to $q$ propagator, as indicated in (1), 
a lattice formulation shows that in fact the symplectic
structure is {\em not} involved; rather, there is always one more $p$
integration than $q$ integration, and that these measures appear
 separately and they 
involve a configuration or momentum space {\em metric}, respectively. 
 Extended in an
invariant way to phase space it suggests that we seek a meaning of the formal
 expression 
of the functional measure ${\cal D}q {\cal D} p$ through  
the introduction 
of a {\sl metric structure} on $M$. It is our opinion that this phase space
metric structure represents the appropriate shadow of quantization
and in some way should replace the notion of polarization 
inside the Geometric Quantization scheme.

 A geometrical quantization procedure moving along these lines has 
 in fact been proposed a few years ago by one of us, J.\ R.\ Klauder 
\cite{Kl1}. In that context the purpose of the (Riemannian) metric 
is to provide an adequate geometrical structure on phase space to 
support Brownian motion which is used to give a continuous-time 
regularization of the  formal expression of the phase-space path integral 
\be
{ K}(p'',q'',t'';p',q',t')\stackrel{\mbox{\scriptsize def}}{=}
       \lim_{\nu\rightarrow\infty} { N}_\nu
       \int
       e^{{i\over\hbar}\int [p_\mu{\dot q}^{\mu}-h(q,p)] dt}\,
       d\mu^\nu_W
\label{KP}
\ee
where $d\mu^\nu_W$ denotes a Wiener measure on $M$ constructed by means 
of the Riemannian metric $g_{ij}$, $\nu$ the Brownian diffusion constant 
and ${ N}_\nu$ an appropriate, and well defined, $\nu$-dependent normalization 
constant. In contrast to the situation depicted in (1) it should be noted that
the Wiener measure on phase space with its pinning of paths at both the initial
and final times leads automatically to an expression depending on $p'',q''$ as
well as $p',q'$. 
 For particular classes of metrics it is possible to demonstrate that 
the propagator (\ref{KP}) with $h(q,p)=0$ behaves as the projector on 
the set of  physical states \cite{Kl1,K&O,MOT,Ma1,AKL,A&K},
while as soon as we consider $h(q,p)\neq0$ we are sure that the states
of the system evolve within the selected subspace.
For highly symmetrical phase spaces, admitting a K\"ahler or a conformal 
K\"ahler structure, the kinematical scheme [$h(q,p)=0$] reproduces the same 
results as the introduction of a complex polarization.
In some sense, therefore, the introduction of a Riemannian metric 
on the phase space $M$  includes the idea of polarization, and, in addition,
remains compatible with the introduction of dynamics. 


Having motivated the phase space metric on mathematical grounds,
we would like to present a different and rather unconventional 
approach to the problem.
 We propose to regard the phase space metric structure as a
concrete physical object---to be considered  as fundamental 
as the symplectic structure and not as an artificial 
regulator (see also \cite{Kl2}). Then we suggest an $\hbar$-dependent 
modification of the laws of dynamics making the quantization problem
into a rather trivial one. In our picture dynamics appears in an 
 interesting way from the competition between the metric and symplectic
phase space structures in close analogy with the guiding center
motion of a charged particle on a plane in an {\em inhomogeneous} magnetic field
\cite{Nor,Gar,Wit,Lj1,Lj2,Ma2}.
 This analogy is actually very useful in picturing both the motion 
of the system and the dynamical mechanism that provides the selection of 
the right set of physical quantum states. 

In section 2 we focus on classical dynamics. After briefly reviewing 
the coordinate-free formulation of classical dynamics---based on symplectic
geometry---we introduce a metric structure on the phase space and  
we illustrate how Hamiltonian mechanics may be described as the adiabatic 
limit of a fully-geometrical phase-space-theory. We also discuss the 
analogy of our model with the motion of a charged particle 
on a manifold in an inhomogeneous magnetic field. As a concrete
example the harmonic oscillator problem is worked out in some detail.
 The problem of quantization is faced in section 3. We start again by
briefly reviewing the mathematical tools necessary for the construction 
of a coordinate-free quantization procedure, 
and, by considering the ``magnetic analogy'', we illustrate 
how this mathematical background is natural in our formulation. 
We then proceed to discuss the quantization of our `free' phase space theory.

Throughout this paper we employ the convention that a sum over repeated 
indices is implied. Phase space coordinates are denoted in a compact manner, 
and they have the dimension of the square root of an action.

\section[]{Hamiltonian Dynamics as `Free' Dynamics} 

\subsection{Kinematics and Symplectic Geometry}

 In the language of modern differential geometry the phase space of an 
$n$ degree of freedom Hamiltonian system is described by
a $2n$-dimensional manifold $M$ equipped with a closed, nondegenerate
two-form, the {\sl symplectic form}\footnote{Throughout this paper 
we shall denote forms and tensors by means of their local components
in a given coordinate frame, e.g.\ $\omega=\omega_{ij}\, d\xi^i\wedge d\xi^j$.} 
$\omega_{ij}$ \cite{A&M,Arn}. This 
geometrical structure, in fact, represents all that is necessary to take 
into account the  kinematical properties of the system, the symplectic 
form being equivalent to the assignment of a Poisson bracket structure
on the phase space. 
  Introducing local coordinates $\xi=(\xi^i; i=1,...,2n)$ on $M$
the components of the symplectic two-form are interpreted as (minus) the 
Lagrange brackets between the phase space coordinates $\lLb\xi^i,\xi^j\rLb=
\omega_{ji}$, so that the fundamental Poisson brackets may  be obtained 
as 
\be
\{\xi^i,\xi^j\}=\,\ob^{ji},
\label{fPb}
\ee
$\ob^{ij}$ being the antisymmetric two-tensor defined in every coordinate 
system by the well-known relation between Lagrange and Poisson brackets, 
$\omega_{ik}\ob^{jk}=\delta_i^j$. 
This  completely characterizes the canonical structure of the system, 
that is the kinematics. 
 
 The description of  dynamics, on the other hand, requires the 
specification of a smooth function on $M$, the {\sl Hamiltonian} $h(\xi)$, 
an object which is not related to any geometrical feature of the 
phase space.
Representing the symplectic two-form by means of the canonical one-form 
$\theta_i$,  $\omega_{ij}=\de_i\theta_j-\de_j\theta_i$, a very convenient 
way to assign dynamics is by means of Hamilton's variational principle
\be
\delta\int\left(\theta_i {\dot \xi^i}
                 - h(\xi)\right)dt=0.
\label{Hvp}
\ee
For a general phase space, $\theta_i$ may be defined only locally and up 
to the gradient of an arbitrary function of $\xi$, $\theta_i\rightarrow
\theta_i+\de_iG$, an arbitrariness which does not affect the results 
of the theory.

\subsubsection*{Canonical Coordinates}

 This phase-space-covariant formulation of mechanics
assumes a more familiar look once {\sl canonical coordinates} are introduced. 
A theorem of Darboux asserts that it is possible to find local 
coordinates such that $\omega_{ij}$ as well as $\ob^{ij}$ reduce to the 
standard form
\be
 \omega_{ij} = \ob^{ij} = \pmatrix{ 0 & -I \cr
                                    I & 0   },
\label{canonical}
\ee
where $I$ represents the $n$-dimensional identity matrix. 
Denoting phase space coordinates by $\xi=(q^1,...,q^n,p_1,...p_n)$, the 
fundamental Poisson brackets (\ref{fPb}) assume the canonical form 
\be
& &\{q^\mu,q^\nu\}=0, \nn          \\
& &\{q^\mu,p_\nu\}=\delta^\mu_\nu, \\
& &\{p_\mu,p_\nu\}=0, \nn
\ee
$\mu,\nu=1,...,n$. Up to the gradient of an arbitrary function of $\xi$, 
the canonical one-form may be chosen as $\theta_i=(p_1,...,p_n,0,...,0)$
so that (\ref{Hvp}) reduces to the standard expression
\be
\delta\int\left(p_\mu{\dot q^\mu}-h(q,p)\right)dt=0. 
\ee
Darboux's coordinates are therefore to  be identified with canonical 
coordinates. 
  In the rest of this paper we suppose that the phase space $M$ is
parametrized by means of canonical coordinate frames. Nevertheless,
in order to simplify the notation and to express our result in a 
phase-space-covariant manner, we continue to denote phase space 
coordinates by means of the single variable $\xi=(q^1,...,q^n,
p_1,...,p_n)$.

\subsection{Dynamics and Metric Geometry}

We now come to the heart of our analysis. 
The global formulation of Hamiltonian mechanics makes it
clear that whereas the kinematical properties of a dynamical system are 
completely taken into account by a geometrical structure,
the symplectic form $\omega_{ij}$, dynamics is described 
by means of a non-geometrical object, the Hamiltonian $h(\xi)$.
 It is the purpose of this section to demonstrate that the  dynamical 
properties of a Hamiltonian system may be understood as  consequence 
of a second geometrical structure on the phase space, a 
{\sl metric} $g_{ij}$. To be more precise we claim that

\vskip0.3truecm
{\em Introducing a  metric $\mu g_{ij}(\xi)$ on the 
phase space $M$ of a Hamiltonian system ($\mu^{1/2}$ being a parameter 
in which the scale of the phase space line-element $ds$ is reabsorbed) 
in the limit of very small values of $\mu$, the variational principle
\be
\delta\int\left({1\over2}\mu g_{ij}{\dot\xi^i}{\dot\xi^j}+
                   \theta_i {\dot \xi^i}\right)dt=0
\label{mHvp}
\ee 
produces the same dynamics as Hamilton's variational principle (\ref{Hvp}),
provided that in any canonical coordinate frame the metric determinant 
$g(\xi)$ satisfies the condition
\be
g(\xi)=h^{-2n}(\xi).
\label{condition}
\ee}
\vskip0.3truecm

\noindent
At first sight, this statement may sound quite strange, the 
replacement of the Hamiltonian $h(\xi)$ with the kinetic-energy-like term 
${1\over2}\mu g_{ij}{\dot\xi^i}{\dot\xi^j}$ making the original
$n$ degree of freedom Hamiltonian theory into a $2n$ degree of
freedom Lagrangian theory. 
The variational principle (\ref{mHvp}) is in fact formally equivalent to 
that describing  the free motion of a particle of mass $\mu$ {\sl (the 
``surface-scale'' of the phase-space)} on a metric manifold $M$ 
{\sl (the phase-space endowed with the metric structure $g_{ij}$)} 
coupled with a kind of universal magnetic field {\sl (the canonical 
two-form $\omega_{ij}$)} \cite{K&O}. 
 This magnetic analogy is actually quite useful in understanding the 
very small $\mu$ regime of the theory, and illustrates 
the mechanism producing the effective removal of the redundant
degrees of freedom of the system.
Before proving our statement let us therefore offer a few additional
words about it.

\subsubsection*{A Magnetic Analogy}

In order to visualize the problem in a very simple case let us 
consider a particle of mass $m$ and charge $e$ moving in a plane
under the influence of a magnetic field of magnitude $B$ normal to the 
plane.  In our analogy the plane represents the phase space of a 
one-dimensional Hamiltonian system whereas the magnetic field its
symplectic structure (see also, in a slightly different context
\cite{Ma1,AKL}).
The limit of a very small  mass corresponds
to that of a very strong magnetic field or, equivalently, to that of a
weakly-inhomogeneous magnetic field.  
The {\em phase-space motion of a dynamical system} will therefore be 
assimilated  into the  {\em adiabatic motion of a charged particle in an 
external magnetic field}. 
 We can learn much about this subject in the literature. The problem, often
referred to as {\sl guiding center motion}, is in fact of primary interest in 
plasma physics and has been treated over the years by many authors from 
 various points of view. An excellent review of the physical principles
may be found in the book of T.\ G.\ Northrop
\cite{Nor}. In view of our interest in the canonical structure of the 
problem, we also refer to the works of C.\ S.\ Gardner \cite{Gar}, 
E.\ Witten \cite{Wit}, R.\ G.\ Littlejohn \cite{Lj1,Lj2} and P.\ Maraner
\cite{Ma2}.

As long as we consider a 
homogeneous magnetic field the particle follows a circular orbit of radius 
$r_B={mc\over eB}|{\vec v}|$ the center of which remains motionless. However, 
as soon as we introduce a weak inhomogeneity the center of the orbit starts 
moving, drifting slowly in the plane. The situation may be described inside 
the canonical formalism by introducing two pairs of canonical variables, 
the {\sl adiabatic kinematical momenta} and the {\sl adiabatic guiding center 
coordinates}. The former takes into account the fast rotation of 
the particle, whereas the latter the slow drift of the center of the orbit. 
 We shall identify the guiding center motion with the phase space motion of 
the dynamical system and the fast rotation of the particle with the redundant
degrees of freedom.
The limit of a very small mass $m\rightarrow 0$, or, equivalently, of a very 
strong magnetic field $B\rightarrow\infty$, induces the circular orbit 
to collapse into a point so that only the guiding center motion remains  
detectable.  The limit of small masses effectively removes the redundant
degree of freedom from the theory simply because it lowers the degree of the
classical equations of motion. 

\subsubsection*{Phase Space Motion in a Universal Magnetic Field}

 The standard analysis of guiding center motion deals with a Euclidean 
configuration space and an inhomogeneous magnetic field. 
For our consideration, we are interested in a possibly more general 
situation in which the metric structure may also vary from point to point. 
 Fortunately, the qualitative picture of the system does not change
since all that matters is
the way in which the magnetic field varies in the given geometry. 
 To be more concrete let us consider the Lagrangian
 $\L(\xi,{\dot\xi})={1\over2}\mu g_{ij}{\dot\xi^i}
{\dot\xi^j}+\theta_i{\dot\xi^i}$. Introducing the canonical momenta $p^\xi_i=
\de\L/\de{\dot\xi^i}$, we consider the corresponding Hamiltonian 
\be
\H(\xi,p^\xi)={1\over2\mu}g^{ij}(\xi)
              \left(p^\xi_i-\theta_i\right)
              \left(p^\xi_j-\theta_j\right),
\label{psH}
\ee
$g^{ij}$ denoting the inverse of the metric tensor.
 It is important not to confuse this extended Hamiltonian theory
 with the original 
Hamiltonian  theory. We are no longer dealing with the phase space $M$, 
 which now appears as the configuration space of our extended system, but with 
its cotangent  bundle $T^*M$ parametrized by the ``positions'' $\xi$ and the 
``momenta''  $p^\xi$ \cite{A&M,Arn}.
In order to avoid any confusion between the original $n$ degrees of freedom 
Hamiltonian system and our extended $2n$ degrees of freedom Hamiltonian system
we shall  denote the Poisson brackets on $T^*M$ by $\lPb\mbox{F},\mbox{G}\rPb=
{\de\mbox{\scriptsize{F}}\over\de\xi^i}
{\de\mbox{\scriptsize{G}}\over\de p^\xi_i }-
{\de\mbox{\scriptsize{F}}\over\de\xi^i}
{\de\mbox{\scriptsize{G}}\over\de p^\xi_i}$.

\subsubsection*{Kinematical Momenta and Guiding Center Coordinates} 

 Let us  proceed by observing that the form of the Hamiltonian (\ref{psH}) 
may be simplified considerably by first replacing the
 canonical momenta $p^\xi_i$ 
with the gauge covariant {\sl kinematical momenta} 
\be
  \Pi_i={1\over\mu^{1/2}}\left(p^\xi_i-\theta_i\right),
\ee
$i=1,...,2n$. Up to a scale factor $\Pi_\nu$ and $\Pi_{n+\nu}$, $\nu=1,...,n$,
behave as conjugate variables so that (\ref{psH}) becomes the Hamiltonian
of an $n$-dimensional harmonic oscillator with masses and frequencies
depending on $\xi$. Since the Poisson brackets between the $\Pi_i$s and the 
$\xi^i$s are in general different from zero, $\lPb\xi,\Pi\rPb\neq 0$, we are 
led to further adapt our phase space variables by introducing the {\sl guiding 
center coordinates}
\be
  X^i= \xi^i +\mu^{1/2}\ob^{ij}\Pi_j,
\ee
$i=1,...,2n$. In our magnetic analogy  the $\Pi_i$s describe the fast 
rotation of the particle around the guiding center, whereas the $X^i$s 
take into account the slow drift of the center of the orbit.
The new set of variables fulfills the Poisson bracket relations
\be
& &\BlPb\Pi_i,\Pi_j\BrPb =\,  \mu^{-1}\ \omega_{ij},    \nn\\
& &\BlPb\Pi_i,X^j\BrPb =\,  0,                \label{ccr}\\
& &\BlPb X^i,X^j\BrPb =\, \ob^{ji},             \nn
\ee
so that the guiding center coordinates and kinematical momenta may be recognized
as a new set of canonical variables (cf. expressions (\ref{fPb}) and 
(\ref{canonical})).
 The presence of the scale factor $\mu^{-1}$ allows us to identify the
$\Pi_i$s 
and the  $X^i$s as describing respectively {\sl fast} and {\sl slow} degrees 
of freedom of the system \cite{Ma2}.
 Rewriting the Hamiltonian (\ref{psH}) in terms of the new variables and 
expanding
in the small parameter $\mu^{1/2}$ we find that
\be
\H(X,\Pi)={1\over2}g^{ij}(X)\Pi_i\Pi_j + {\cal O}(\mu^{1/2}).
\label{cH}
\ee
 The relevant term of the expansion looks again like an $n$-dimensional 
harmonic oscillator in the fast variables $\Pi_i$s the parameters
depending this time only on the slow variables $X^i$s. 

\subsubsection*{A second canonical transformation}

 The dynamics of fast and slow degrees of freedom may be separated,
up to terms of order $\mu^{1/2}$, by performing a second canonical transformation.
For this task we decompose the inverse metric $g^{ij}(X)$ as $g^{ij}(X)=
g^{-1/2n}(X)\gamma^{ij}(X)$, $g(X)$ being the determinant of the 
metric and $\gamma^{ij}(X)$ a point-dependent matrix with  determinant one. 
We further represent $\gamma^{ij}(X)$ by means of $2n$-beins 
as $\gamma^{ij}(X)=\delta^{kl}\t_k^i(X)\t_l^j(X)$. Making use of the condition
(\ref{condition}) the inverse metric may thus be written as
\be
g^{ij}(X)=h(X)\ \delta^{kl}\t_k^i(X)\t_l^j(X),
\ee
where $h(X)$ is  the Hamiltonian of our original system.
 Denoting by $\tau^i_j(X)$ the logarithm of the $2n$-bein $\t^i_j(X)$,
$\tau(X)=\ln\t(X)$, we perform a canonical transformation generated 
by the  function $\Lambda(X,\Pi)= {1\over2}\tau^i_k(X)\ob^{kj}
\Pi_i\Pi_j$, the infinitesimal parameter being identified with $\mu$. 
 The variables produced by the transformation
again fulfill the Poisson brackets relations (\ref{ccr}) so that 
the new phase space coordinates are again separated into the two canonical 
subsets $\{X'\}$ and $\{\Pi'\}$, ${X'}^\nu$ being conjugate to ${X'}^{n+\nu}$ 
and $\Pi'_\nu$ to  $\Pi'_{n+\nu}$, $\nu=1,...,n$. Up to terms of order $\mu$ it
 follows that
\be
\left\{
\begin{array}{rcl}
{X'}^i&=&X^i+{\cal O}(\mu) \\
{\Pi'}_i&=&\t_i^k(X)\Pi_k+{\cal O}(\mu)
\end{array}
\right. .
\ee
In terms of the new variables the Hamiltonian (\ref{psH}) separates into
a product of a function of the $X'^i$s  times a function of the $\Pi'_i$s
\be
\H(X',\Pi')= h(X')\ J +{\cal O}(\mu^{1/2}),
\label{H}
\ee
$J={1\over2}\sum_i{\Pi'_i}^2$ representing the Hamiltonian of an
$n$-dimensional harmonic oscillator. 

\subsubsection*{(Effective) Hamiltonian Dynamics}

 Disregarding higher order terms, the mechanics of the $X'^i$s completely 
separates from that of the $\Pi'_i$s. The $X'^i$s describe an $n$ degree
of freedom Hamiltonian system the phase space of which may be identified with
 $M$ 
and whose dynamics is characterized by the Hamiltonian $h$, namely our 
original Hamiltonian system. The $\Pi'_i$s, on the other hand, describe an
$n$ dimensional harmonic oscillator performing, for fixed energy, vibrations 
of amplitude $\mu$ and frequency $\mu^{-1}$. 
By decreasing the value of $\mu$ the orbits of our extended system collapse 
therefore into ones of the original Hamiltonian system, the presence 
of the redundant variables $\Pi'$ becoming increasingly irrelevant.

 We would like to stress that we are not considering the limit procedure
$\mu\rightarrow 0$ in a rigorous mathematical sense. In our present viewpoint 
$\mu^{1/2}$ represents a very small but {\em finite} parameter, capable of a 
concrete physical interpretation. It represents the phase-space-length-scale 
over which the universal magnetic field represented by the symplectic
two-form $\omega_{ij}$ may be considered as homogeneous. On the other hand, it
is the inhomogeneities on larger scales that produce dynamics.

\subsection[]{The Harmonic Oscillator Problem as a Simple \\
              Example of the Method}

For the sake of completeness let us write down explicitly the 
geometrical equations driving our dynamical theory. 
Consider an $n$-degree of freedom system described by the (positive 
definite) Hamiltonian $h(\xi)$. On the phase space $M$ we introduce 
the metric 
\be
g_{ij}(\xi)={1\over h(\xi)}\gamma_{ij}(\xi),
\label{dec}
\ee
$\gamma_{ij}$ being a point dependent
$2n$ by $2n$ matrix with determinant one. The choice of $\gamma_{ij}$ is 
obviously related to the topological features of the phase space. As long 
as we are interested in the adiabatic regime its explicit form does not 
play any role and its choice is purely a matter of convenience.
 For a flat topology 
we may  choose the Kronecker delta, $\gamma_{ij}=\delta_{ij}$, while 
non trivial topologies generally require more complicated expressions.
The equations of motion follow from the Lagrangian 
$\L(\xi,{\dot\xi})$ as
\be
\ddot{\xi}^k+\Gamma^{k}_{ij}\dot{\xi}^i\dot{\xi}^j
         ={1\over\mu}g^{ki}\omega_{ij}\dot{\xi}^j\
\label{feqs}
\ee
$k=1,...,2n$ and $\Gamma^{k}_{ij}=g^{kl}(\de_ig_{lj}+\de_jg_{il}-
\de_lg_{ij})/2$ denoting the Christoffel symbols relative to the 
connection induced on $M$ by $g_{ij}$. Aside from the magnetic 
term on the right hand side, these correspond to the geodesic
equations for a free motion on $M$. Nevertheless, it has to be 
pointed out that the presence of the Lorentz like term ${1\over\mu}
g^{ki}\omega_{ij}\dot{\xi}^j$  can drastically modify the behaviour of 
the system, even for large  values of $\mu$.
By decreasing the value of $\mu$ further, the trajectories of our system
tightly wrap around the ones of the original Hamiltonian system,
becoming physically indistinguishable from these for very small values 
of $\mu$. In order to illustrate these features in a concrete example let us
discuss in some detail the harmonic oscillator problem.

\subsubsection*{The harmonic oscillator}

 Consider a one-dimensional harmonic oscillator described by the Hamiltonian
$h(p,q)={1\over2}(p^2+q^2)$. The topology of the phase space being trivial 
we choose the metric tensor $g_{ij}(p,q)=2\delta_{ij}/(p^2+q^2)$. 
 This make the phase plane into an infinite cylinder, the extremities of which
have to be identified with the inaccessible point zero\footnote{The origin may 
be made into an accessible point for the system by adding a positive constant 
to the Hamiltonian and hence to the conformal factor of the metric. This 
modifies the geometry of the phase plane (it is no longer flat) 
but not the adiabatic regime of the dynamics. Moreover, our description 
of dynamics is in some sense fuzzy. We consider phase space points
infinitesimally 
close to each other as indistinguishable physical states so that the loss
of a single phase space point does not constitute a serious problem.} 
and the point at infinity. To make this explicit we introduce  
non-canonical cylindrical coordinates $(\rho,\phi)$ related to $(p,q)$ by the 
transformation $q=\mu^{1/2}\mbox{e}^{-\rho}\sin\phi$,
$p=\mu^{1/2}\mbox{e}^{-\rho} \cos\phi$. 
Choosing the symmetric gauge for the canonical one-form $\theta_i$, the 
phase space Lagrangian of the system reads
\be
 \L(\rho,\phi,\dot{\rho},\dot{\phi})= 
            \mu\,\left(\dot{\rho}^2+\dot{\phi}^2\right)
          +{\mu\over2}\,\mbox{e}^{-2\rho}\dot{\phi}
\ee 
making clear the formal analogy of our system with a particle moving on a 
cylinder in an orthogonal magnetic field of magnitude 
$B(\rho)\simeq -\mbox{e}^{-2\rho}$. 
 The presence of the magnetic term  makes the region $\rho= -\infty$ 
inaccessible, dramatically modifying the free trajectories of the system.
The geodesics  on the cylinder are in fact represented by circles of constant 
$\rho$ and  helices escaping toward both extremities with constant velocity. In 
the phase plane picture of the cylinder these trajectories are represented 
respectively by circles, $r(t)\equiv\sqrt{p^2(t)+q^2(t)}=const$, and by spirals 
collapsing onto the origin, $r(t)\sim r_0\,\mbox{e}^{-kt}$, or escaping to 
infinity, $r(t)\sim r_0\,\mbox{e}^{kt}$. For every value of $\mu$ the magnetic
force removes the  trajectories escaping to infinity by confining the
motion to a neighbourhood of the origin.

\begin{figure}[t]
\centerline{\mbox{\epsfig{file=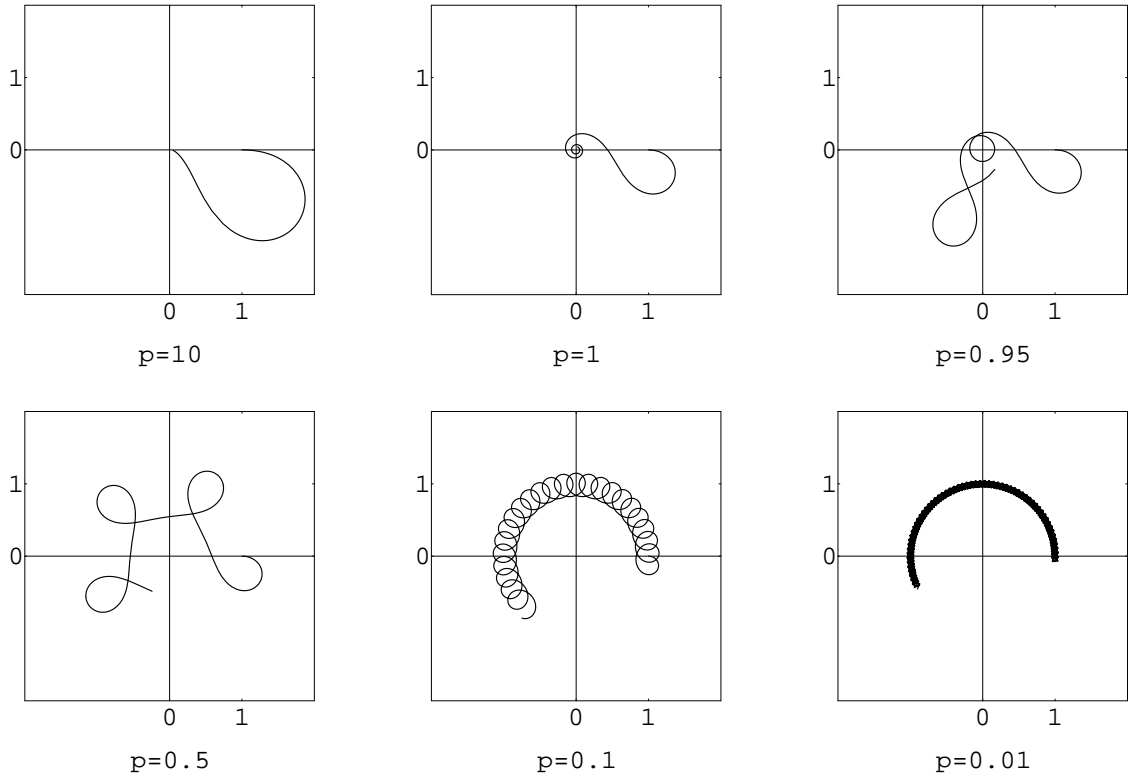}}}
\caption{Phase space motion of the representative point of the system for 
different values of the parameter $\p={\mu{\cal E}\over l^2}$. The system
is initially in the point $(1,0)$. The ``energy'' ${\cal E}$ and 
the ``angular momentum'' $l$ are fixed to the values $1$ and  $1\over4$ 
respectively.}
\label{fig1}
\end{figure}

In order to proceed to the solution of the dynamical problem we consider
the two integrals of motion of the system, the analogues of angular momentum 
and energy for the equivalent particle moving on the cylinder,
\be
\mu\,\dot{\phi}+{1\over4}\,r^2=l,\label{am}\\
\mu\,{\dot{r}^2\over r^2}+\mu\,\dot{\phi}^2={\cal E}.\label{energy}
\ee
By eliminating $\dot{\phi}$ in (\ref{energy}) by means of (\ref{am}) (we
suppose $l\neq 0$) we see that the motion in the $\rho$ 
direction takes place in a Morse potential. Introducing the variable 
$\zeta={\mu \over4l}e^{-2\rho}-1$ the quadrature of the problem is then 
reduced to the evaluation of the integral
\be
t-t_0=\pm{\mu\over 2l}\int_{\zeta_0}^{\zeta}
       {d\zeta\over (\zeta+1)\sqrt{{\mu{\cal E}\over\l^2}-\zeta^2}}.
\label{integral}
\ee
This yields
\begin{equation}
r^2(t)=\left\{
\begin{array}{ll}
\ds2l\,(\p-1)\over\ds
      e^{\pm 2l\sqrt{\pp-1}(t-t_0)/\mu}
 +\p\,e^{\mp 2l\sqrt{\pp-1}(t-t_0)/\mu}-2
   & \mbox{for}\;\p>1 \\
   & \\
\ds8l\over\ds
 1+4l^2(t-t_0)^2/\mu^2
  & \mbox{for}\;\p=1 \\
  &  \\
\ds4l\,(1-\p)\over\ds 
 1\pm \p^{1/2} \sin\left[2l(1-\p)^{1/2}(t-t_0)/\mu\right]& 
  \mbox{for}\;\p<1
\end{array}
\right.
\end{equation}
The behaviour of the system depends on the parameter 
$\p=\mu{\cal E}/l^2$, its value being greater, equal or lesser 
than one producing three different dynamical regimes. The trajectories 
with $\p>1$ correspond to unbound states of the Morse potential. 
In the phase plane picture of the system the representative point 
falls onto the origin with the exponential law $r(t)\sim e^{-t}$
(Fig.\ref{fig1}, $\p=10$). For $\p=1$ the ``energy'' of the system 
equals the asymptotic limit of the Morse potential so that the motion
is again unbounded. The phase space trajectories again fall onto the 
origin but with the power law $r(t)\sim 1/t$ (Fig.\ref{fig1}, $\p=1$).  
Finally, for $\p<1$, we obtain the bound states of the Morse potential. 
The representative point of the system neither falls onto the phase plane 
origin nor escapes to infinity.

Whereas for $\p\geq1$ the trajectories of the system share a quite
simple form, for values of $\p$ very close to one from below the
representative point of the system makes rather unusual curves on the 
phase plane trying  to fall onto the origin but returning over and over 
to a neighbourhood of the starting point (Fig.\ref{fig1}, $\p=0.95$).
For fixed values of ${\cal E}$ and $l$ the adiabatic limit of the theory
is reached for very small values of $\mu$.
 By  decreasing $\mu$, in fact, the oscillations of $r(t)$ and
also of $\phi(t)$ are strongly damped so that for very small $\mu$
the system follows a thick spiral of very small radius wrapping around 
a circle, that is a phase space trajectory of the harmonic oscillator
(Fig.\ref{fig1}, $\p=0.5,\p=0.1,\p=0.01$). This is exactly  the adiabatic 
behaviour we have predicted in general terms.

\section{`Free' Quantum Dynamics} 

\subsection{The Geometrical Background of Quantization}

 Some thirty years ago the problem of quantizing a general 
Hamiltonian system has been seriously faced for the first time
in the so called Geometric Quantization scheme of B.\ Kostant,
A.\ Kirillov  and J.\ M.\ Souriau \cite{Kos,Kir,Sou,Woo}. 
Geometric Quantization should not  be considered by the same
standard as the several physics-generated 
quantization procedures that have been proposed over 
the years. Rather, it should be regarded as an analysis of 
the various structures needed for the quantization of a classical system,
providing the proper mathematical background and the right mathematical 
tools necessary to analyze the issues surrounding quantization.
On the other hand, Geometric Quantization  lacks physical 
intuition and, as a matter of fact, it has succeeded more in pointing out 
the formal difficulties involved in the quantization procedure 
than in providing their solution. 
 Though in what follows we will make only an implicit use of the 
abstract tools introduced by Geometric Quantization, this language exactly 
corresponds to the one to be employed in the description of a charged quantum 
particle in a non-trivial topology, that is, taking into account the 
magnetic analogy we discussed in the previous sections, in our 
dynamical theory (see also \cite{K&O}). We find it worthwhile,
therefore, to briefly recall the salient features of the 
construction.

In discussing a field theory like quantum mechanics in a non-trivial 
topological context it is necessary to pay attention
in treating  global features \cite{W&Y}. Although everything 
should make sense globally  not every object appearing in the theory 
is capable of a global definition. As a relevant example, 
once a phase space with a non-trivial topology is considered the 
canonical one-form $\theta_i$ is only locally defined (like 
the vector potential of an Aharonov-Bohm magnetic field). This 
quantity, appearing directly in the Hamiltonian action,  
 forces the wave functions of the system to share the same
undesirable  feature. The problem, nevertheless, does not concern 
the theory as a whole but only its local representation and 
an appropriate language to deal with the situation has to be introduced. 
This is fiber bundle theory \cite{Woo}. We have in some sense to be content 
with a piecewise representation of the theory making sure that when moving 
from one  local representation to another,  
everything makes sense globally. In constructing a coordinate-free
quantization procedure, therefore, we have to take the symplectic 
two-form as the {\sl curvature form} of an appropriate {\sl line bundle}
over the phase space $M$. The (only locally defined) canonical 
one-form appears then as the corresponding {\sl connection form} while 
the wave functions of the system acquire a global meaning as {\sl sections} 
of the line bundle \cite{Kos}.
 The practical results of this elegant construction \cite{Kos}---which 
is the only way to give a global meaning to the world ``quantization''---is
that the so constructed Hilbert space appears to be too 
large and some additional
structure must be picked out on the phase space $M$ in order to reduce
its dimension.
 This brings us back to the introductory section and to the discussion
concerning real/complex polarizations and phase space metric structures.
For details we refer to the original works quoted above.
An interesting approach, similar in  many respects to that of polarization, 
has also being recently developed by E.\ Gozzi \cite{Goz}.

\subsection{Quantizing `Free' Dynamics}

 The task of giving a fully geometrical picture of the dynamical 
mechanism leading to the set of physical states 
for a quantum system is the main motivation which has brought us 
to a description of standard Hamiltonian mechanics as the adiabatic limit 
of a fully geometrical phase-space-theory. 
Once classical dynamics is re-expressed in terms of the variational 
principle (\ref{mHvp}), the task of quantizing the classical 
system is reduced to standard procedures.

\subsubsection*{Path Integral Approach}

The basic features of this approach
 may be seen immediately by writing down the formal phase space 
expression of the propagator
\be
 K(\xi'',t'';\xi',t')=\int 
e^{i\int\left(
{1\over2}g_{ij}{\dot\xi^i}{\dot\xi^j}+{1\over\mu}\theta_i{\dot\xi^i}
\right)\,dt} {\cal D}\xi;
\label{ourP}
\ee
the presence of the kinetic-energy-like term ${1\over2}\mu g_{ij}
{\dot\xi^i}{\dot\xi^j}$ in the phase space action enables one to 
give a precise---although not unique since ordering ambiguities are still
present---meaning to this expression by means of an imaginary time 
continuation and a Wiener measure on $M$, exactly  as in Klauder's 
quantization scheme \cite{Kl1}. Nevertheless, in the present
context we need not perform any limiting procedure to remove any 
regulator, the phase space metric playing now  an essential dynamical 
role in the theory.

\subsubsection*{Hamiltonian Approach}

 An alternative way to look at the standard nature of quantization
in our scheme is to think of the magnetic analogy. The problem is
equivalent to that of quantizing a particle moving on a
metric manifold $M$ in the universal magnetic field $\omega_{ij}$. 
 As sketched in the previous section, in discussing the motion of 
a charged quantum particle in an external magnetic field 
in a non-trivial topology, it is necessary to treat global 
properties very carefully. On the other hand, the problem is 
a fairly standard one. From the work of T.\ T.\ Wu and C.\ N.\ Yang 
on the geometrical setting of Dirac's monopole theory \cite{W&Y} 
we learn that the magnetic field and  vector potential (the canonical 
two-form and one-form, in our context) have to be 
considered respectively as the curvature two-form and the connection 
one-form of an appropriate line bundle over the configuration space (the
phase space $M$, in our context), while the states of the system have 
to be identified with sections of this line bundle (see also \cite{Ma1,D&V}). 
The whole  apparatus of geometric quantization reappears therefore in a 
very natural and necessary manner.
 The Hamiltonian operator associated to the propagator (\ref{ourP}) 
is also capable of a {\em global} definition in terms of the Laplacian
over the considered line bundle and, eventually, invariant counterterms 
constructed  by means of the phase space metric and symplectic structures.
In any coordinate frame the quantum Hamiltonian will appear as
\be
\H={1\over2g^{1/2}(\xi)}\,\Pi_i\, g^{ij}(\xi)\,g^{1/2}\,(\xi)\Pi_j
   +\mu\,{\cal I}_1 +\mu^2\,{\cal I}_2 + ...,
\label{ourH}
\ee
where we have introduced the kinematical momenta $\Pi_i=-i\mu^{1/2}\de_i
-\theta_i/\mu^{1/2}$ and ${\cal I}_1$, ${\cal I}_2$, etc.\ , are 
``optional'' invariants whose presence reflects the ordering ambiguities
inherent in the quantization procedure. As an example ${\cal I}_1$ may 
contain a term proportional to the phase space scalar curvature $R$ \cite{DeW},
but also other invariants with the right dimension constructed from 
the covariant derivatives of $\omega_{ij}$ are possible. 
These invariants produce ${\cal O}(\hbar^2)$ corrections 
to the spectrum of the system, effects which are generally small.
For the moment we do not care to make any 
particular choice of them; a quite natural choice will appear later.

We observe that the Hamiltonian (\ref{ourH}) acts on wave functions depending 
on all the phase space coordinates $\xi=(q,p)$ so that at first sight 
it may appear that our theory shares the same difficulties as Kostant's 
prequantization scheme. However, an analysis along the same lines 
 as that in section 2.2 indicates that the system provides, 
{\em by itself}, the means to remove the unphysical
 degrees of freedom, the intuitive 
picture to keep in mind being always that of Fig.\ref{fig1}.

\subsubsection*{Kinematical Momenta and Guiding Center Operators}

In close analogy with our discussion of the classical theory we introduce,
besides the kinematical momenta $\Pi_i$, the guiding center operators
$X^i=\xi^i+\mu^{1/2}\ob^{ij}\Pi_j$ obtaining a new set of observables. 
 In any canonical coordinate frame the local representation of the 
$X$'s and $\Pi$'s as differential 
operators satisfy the canonical commutation relations
\be
& & \Big[\Pi_i,\Pi_j \Big] =\,i\omega_{ij},    \nn\\
& & \Big[\Pi_i, X^j  \Big] =\,0,     \label{cqcr}\\
& & \Big[ X^i,  X^j  \Big] =\,i\mu\, \ob^{ji}.  \nn
\ee
It is nevertheless important to stress that, unless the topology of the phase 
space  $M$ is trivial, these commutation relations hold only {\em locally} ! 
That is to say the $X$'s and $\Pi$'s do not constitute in general a global
representation of the Heisenberg algebra. On the other hand, we are 
not trying to construct a quantization procedure in the standard sense, namely
looking for a correspondence between the algebra of classical and 
quantum observables; all that we are looking for is a global 
definition of the dynamics of the system and (\ref{ourH}) is 
in fact (a local representation of) a globally well defined object.

\subsubsection*{The Adiabatic Expansion}

Replacing $\xi^i$ with $X^i-\mu^{1/2}\ob^{ij}\Pi_j$ in (\ref{ourH}) and
expanding in power of $\mu^{1/2}$ we obtain the quantum analog of equation
(\ref{cH}). As in the classical case the $X$ and $\Pi$ degrees of freedom
may be separated up to terms of order $\mu^{1/2}$ by performing a unitary
transformation generated by the Hermitian operator $\Lambda(X,\Pi)=
{1\over2}\tau^i_k\ob^{kj}\left\{\Pi_i,\Pi_j\right\}$ ($\{\, ,\, \}$ 
denotes anticommutators here). 
By successive suitable unitary transformations it
is also possible to make all the half-integer order terms of the 
perturbative expansion vanish identically,
while the integer order terms may be written as geometric invariants 
evaluated in the $X$'s times powers of the harmonic oscillator 
Hamiltonian  $J={1\over2}\sum_i\Pi_i^2$ constructed by means of 
the $\Pi$'s (the method to be used is a straightforward generalization
of a well-known technique of perturbation theory in classical mechanics 
and has been  developed in \cite{Ma2}). 
 
 Denoting again by $X^i$ and $\Pi_i$ the new ``canonical'' 
operators---fulfilling (\ref{cqcr}) in every canonical coordinate 
frame---the quantum Hamiltonian describing our system takes on the form
\be
\H=h(X)\, J\, +{\cal O}(\mu).
\label{Hend}
\ee
The original Hamiltonian $h(\xi)$ is here evaluated in the set of 
non-commuting operators $X$, an operation involving ordering ambiguities. 
It is on the other hand immediately realized that a different choice 
of ordering modifies only the higher order terms of the expansion, 
terms which are already not uniquely defined in virtue of the 
freedom in the choice of the invariants ${\cal I}_1$, ${\cal I}_2$, etc.\  .

\subsubsection*{(Effective) Quantum Dynamics}

The dynamics of the $2n$  canonically conjugate {\sl slow} variables $X$'s 
separates from that of the {\sl fast} $\Pi$'s. The energy necessary to induce 
a transition in the spectrum of the fast variables being much greater than the
energy scale involved in the slow motion, the system may be considered as
frozen in one of the $J$ eigenstates and 
the effective dynamics pertains only to the evolution of the slow variables. 
In other words, the system, by itself, effectively removes dynamically the 
redundant (physically unobservable) degrees of freedom.
 The higher order terms of the perturbative expansion being  operators
depending on the variables $X$'s---commuting to $i$ times the adiabatic
parameter $\mu$---contribute to the spectrum of the system with 
corrections of order higher than $\mu^2$. Moreover, once the system is 
frozen in an eigenstate of $J$, presumably its ground state, it is possible
in principle to perform a choice of the invariants ${\cal I}_1$, 
${\cal I}_2$, etc.\ , in such a way that the whole adiabatic expansion 
except for the zero order term identically vanishes for that state. The scheme
therefore allows a reproduction of all the ordering prescriptions and even
something more.  

 In concluding this section let us observe what the reader probably 
already suspects. A rapid look at the commutation relations (\ref{cqcr}), 
the Hamiltonian (\ref{Hend}) and even the propagator (\ref{ourP}),
makes clear  that the adiabatic parameter $\mu$ should be 
identified with Planck's constant
\be
\mu\equiv\hbar.
\ee
Hereafter, we shall assume this equality.
In our picture, therefore, Planck's constant assume an intuitive geometrical 
meaning: $\hbar^{1/2}$ is the natural phase-space-length-scale
measuring the inhomogeneity of the universal magnetic field $\omega_{ij}$ 
in the metric $g_{ij}$.

\subsection{One Degree of Freedom Systems}

In order to illustrate in more detail the method and to compare it with
standard quantization procedures in a trivial topological context
we specialize to one degree of freedom systems.
The phase space to be considered is then represented by a two-dimensional 
surface $M$ while---as in the general case---quantum kinematics and dynamics 
are completely characterized by (\ref{ourP})/(\ref{ourH}) once symplectic 
and metric structures are assigned. In every canonical coordinate frame 
$\xi=(q,p)$
\be
\omega_{ij} =               \pmatrix{ 0 & -1 \cr
                                      1 &  0   },
\ \ \ \
g_{ij}      ={1\over h(\xi)}\pmatrix{ \gamma_{11} & \gamma_{12} \cr
                                      \gamma_{12} & \gamma_{22}   },
\label{1df1}
\ee
where the metric has again been factored into the product of a 
function times a point dependent matrix with determinant one 
as in (\ref{dec}). Let us observe that in the case of one degree
of freedom systems this decomposition has a special covariant character. 
The inverse conformal factor $h(\xi)$ corresponds  in fact to
the norm of the symplectic two-form $\omega_{ij}$, $h(\xi)=\sqrt{\omega_{ij} 
\omega^{ij}/2}$. $h(\xi)$ transforms therefore as a scalar while $\gamma_{ij}$
as a symmetric two-tensor.

Suppose now that the topology of the surface $M$ is compatible with 
a flat geometry. This is the case, as an example, of the harmonic oscillator 
discussed in section 2.3 and of most dynamical system usually considered in 
textbooks. Without affecting the adiabatic regime of the theory---that is 
dynamics---it is then possible to choose the tensor $\gamma_{ij}$ in such 
a way that $g_{ij}$ is flat. 
Performing this choice eliminates geometrical complications, the problem
resulting being equivalent (up to boundary conditions) to the motion of a charged
spinless particle in a plane under the influence of a perpendicular
inhomogeneous magnetic field.
To make this explicit we introduce Cartesian (non-canonical) coordinates
$\bar{\xi}=\bar{\xi}(\xi)$. The metric tensor then becomes a  
Kronecker delta while it follows that the symplectic 
two-form is simply multiplied by its norm,
\be
\bar{\omega}_{ij} = h(\bar{\xi})\pmatrix{ 0 & -1 \cr
                                         1 & 0    },
\ \ \ \ 
\bar{g}_{ij}      =                  \pmatrix{ 1 & 0 \cr
                                         0 & 1   }.
\label{1df2}
\ee
The bar indicates that the tensors are to be evaluated in the new coordinates
while $h(\bar{\xi})$ should be interpreted as $h(\xi(\bar{\xi}))$.
In the Cartesian background the Hamiltonian (\ref{ourH}) becomes
\be
\H={1\over2}\delta^{ij}\bar{\Pi}_i\bar{\Pi}_j
   +\hbar\ {\cal I}_1+\hbar^2\ {\cal I}_2+ ...\, ,
\label{H1d}
\ee
$\bar{\Pi}_i={\de \xi^k/\de\bar{\xi}^i}\Pi_k$ denoting the Cartesian
kinematical momenta, and the invariants ${\cal I}_1$, ${\cal I}_2$, etc.\ ,
 are evaluated in $\bar{\xi}$. Obviously $\bar{\Pi}_1$ and $\bar{\Pi}_2$ 
are no longer conjugate variables. The new set of operators 
$\bar{\Pi}$'s and $\bar{\xi}$'s in fact fulfill the commutation relations
\be
& & \Big[\bar{\Pi}_i,\bar{\Pi}_j \Big] =\,i\,h(\bar{\xi})\,\omega_{ij}, 
                                                                          \nn\\
& & \Big[\bar{\Pi}_i,\bar{\xi}^j \Big] =- i\,\hbar^{1/2}\,\delta_i^j,
                                                                 \label{cr1d}\\
& & \Big[\bar{\xi}^i,\bar{\xi}^j \Big] =\,0.  \nn
\ee
The Hamiltonian (\ref{H1d}) together with   (\ref{cr1d}) 
makes explicit the analogy of the problem with the motion of a quantum charged 
particle  in a plane under the influence of the magnetic field $B(\bar{\xi})
=h(\bar{\xi})$ \cite{Ma2}. The adiabatic regime of this theory has been 
recently investigated by one of us, P.\ Maraner, obtaining the explicit 
expression of the first few terms of the adiabatic expansion. Introducing 
in a suitable way {\sl adiabatic kinematical momenta} and 
{\sl adiabatic guiding center operators} the Hamiltonian (\ref{H1d}) 
becomes (see \cite{Ma2} for details)
\be
\H= h\, \bar{J}+
   {\hbar\over4}\left[
              {\triangle h\over h}-
             3{|\nabla h|^2\over h^2}
             \right]\,\bar{J}^2
  +{\hbar\over16}\left[
               {\triangle h\over h}-
               {|\nabla h|^2\over h^2}
               \right]\,
  +\hbar{\cal I}_1 +{\cal O}(\hbar^2),
\ee
where $\bar{J}$ represents the harmonic oscillator Hamiltonian constructed
by means of the adiabatic kinematical momenta and all the scalars are 
evaluated in the adiabatic guiding center operators. 
Freezing the fast variable of the system in its ground state and 
transforming back to the original canonical frame the effective 
Hamiltonian $h^{(eff)}$ describing the slow motion is obtained as
\be
h^{(eff)}= {1\over2}h(X)+\hbar\left[{1\over8}{\triangle h\over h}(X)-
                           {1\over4}{|\nabla h|^2\over h^2}(X)+
                           {\cal I}_1(X)\right] +{\cal O}(\hbar^2),
\ee
$X^i=\xi^i+\hbar^{1/2}\ob^{ij}\Pi_j$, $i=1,2$, again denoting the guiding 
center operators introduced in the previous section. For any arbitrarily
assigned ordering prescription, the choice (compare also \cite{MOT})
\be
{\cal I}_1={1\over4}{|\nabla h|^2\over h^2}-
           {1\over8}{\triangle h\over h}
\ee
makes our quantization scheme  reproduce the standard one up to terms
of order $\hbar^3$. It is  also possible, at least in principle, 
to proceed by choosing all the remaining invariants ${\cal I}_2$, 
${\cal I}_3$, etc.\ , is such a way that the whole perturbative expansion 
except the zero order term vanishes identically. Aside from an inessential
multiplicative factor $1/2$ the {\sl (effective) quantum dynamics} of 
the system is  described by
\be
h^{(eff)}=\,h(Q,P),
\ee
$Q\equiv X^1$ and $P\equiv X^2$ being a pair of conjugate operators, 
$[Q,P]=i\hbar$, and where an  ordering choice has been performed.

\section{Discussion and Speculations}

Starting from the generally well accepted opinion that quantization
involves picking out some {\em additional structure} on the phase space $M$
of a classical system we have speculated on the possibility of describing
classical as well as quantum dynamics by means of a {\sl phase space metric 
structure}. This  produces an ${\cal O}(\hbar)$ modification of the classical
equations of motion reducing at the same time the problem of 
quantizing an arbitrary Hamiltonian system to standard procedures.
 Our analysis nevertheless appears as unconventional. We do {\em\bf not} 
insist, in fact, on a unique correspondence between classical and quantum 
states, observables and evolution equations. All that we care about is 
giving a global definition of quantum dynamics in the Hilbert space of 
square integrable functions on the classical phase space $M$ (see (\ref{ourP}),
(\ref{ourH})). The system then provides {\em by itself}
  the dynamical selection 
of the subspace of physical states. 
Moreover, our scheme does not yield a unique answer to quantization.
Questions connected with ordering  are still present in the 
theory. On the other hand, as long as various physical situations potentially
involve different orderings, it is our opinion that a sensible quantization
scheme should give 
 not {\sl one quantization} but {\sl ``all'' 
possible quantizations} of any given classical system. 

In our view, dynamics appears in a very interesting way as a purely 
geometrical effect, in formal analogy with the guiding center motion 
of a charged particle in a curved background ({\sl the additional metric 
structure}) under the influence of a universal magnetic field ({\sl the 
classical symplectic structure}). 
 In the present paper we have restricted our attention to non-singular
symplectic structures, that is to unconstrained systems. Nevertheless,
there is no problem, at least in principle, in extending our discussion 
to singular symplectic structures since the dynamics is supported by the 
metric. As a very simple but nontrivial example we may consider 
motion in a three dimensional phase space. The symplectic structure 
is then singular and yet we can still picture the behaviour of the system 
by means of the motion of a particle in an ordinary three-dimensional
space under the influence of an arbitrary magnetic field (compare with
section 3.3).
The resulting adiabatic picture \cite{Nor} is that of a system moving 
freely along the field lines (the ``unphysical'' part of dynamics) while 
rapidly rotating around its guiding center (the unobservable 
degree of freedom) and drifting in the directions normal to the field 
(the ``physical'' part of dynamics). 
 The principal obstacle in extracting an explicit form for the 
Hamiltonian describing the effective guiding center motion, namely
the physically relevant part of dynamics, is deeply connected with the 
problem of finding a local Darboux coordinate frame in which the magnetic 
field reduces to the canonical form \cite{Gar}
\be
 \omega_{ij}  = \pmatrix{ 0 & -1 & 0 \cr
                          1 &  0 & 0 \cr
                          0 &  0 & 0 }.
\nonumber
\ee 
 Succeeding in this task is on the other hand equivalent to the so 
called {\sl abelianization of the constraints} representing a
complete separation of the physical and unphysical degrees of freedom, 
and which leads directly to
the solution of the problem. What appears interesting from our point of view
is that, in the study 
of guiding center motion, techniques have been developed to describe
the adiabatic regime of the dynamics without directly appealing to the 
explicit form of the Darboux transformation \cite{Lj3}.
Our scheme appears therefore as promising in dealing 
with the quantization of {\sl constrained systems}. 

 From a more speculative viewpoint other interesting questions may be 
addressed: 

We may wonder, as an example, if the ``unobservable'' degrees 
of freedom represented by the fast rotation of the system around 
its guiding center are capable of a physical interpretation
(that is, if they are observable after all).
A reasonable guess would be that the $SU(n)$ hidden 
symmetry of our dynamics may accomodate the {\sl spin} degrees of 
freedom of a quantum system. To clarify this point one needs to study the 
response of the system to an external magnetic field, which may be 
incorporated into the theory as a local modification of the 
symplectic structure.

 More ambitiously, one may speculate on the possibility that 
the ${\cal O}(\hbar)$ modification of classical mechanics presented 
in Eq.\ref{mHvp} is in some way related to quantum mechanics 
itself---without going through quantization---as the fuzzy 
trajectories of Fig.\ref{fig1} may suggest. Nevertheless, even in 
the solution of the simple harmonic oscillator problem there 
is no trace of quantization and every attempt at constructing a statistical
theory based on a deterministic background must deal with Bell's theorem.

 Finally, one may wonder about the possibility of giving
a dynamical role to our metric, relating phase-space-geometry
to the phase-space-matter-di\-stri\-bu\-tion in a way reminding 
one of general relativity. 

At the moment, however, these points go well beyond our original purpose.

\section*{Acknoledgments}

It is a genuine pleasure to thank Enrico Onofri for many stimulating 
discussions on the geometrical roots of classical and quantum 
mechanics. J.R.K.\ acknowledges many discussion with Robert Alicki
on related questions. P.M.\ is  pleased to acknowledge conversations
with Ennio Gozzi and Roberto De Pietri.

\end{document}